\documentclass[11pt]{article}
\usepackage{acl2015}
\usepackage{times}
\usepackage{url}
\usepackage{latexsym}

\usepackage{graphicx}
\usepackage{multirow}
\usepackage{makecell}
\usepackage{xcolor}
\def\projTitle{GraphScholarBERT}

\title{Combining Language and Graph Models for Semi-structured Information Extraction on the Web}

\author{Zhi Hong \\
  University of Chicago \\
  {\tt hongzhi@uchicago.edu} \\\And
  Kyle Chard, Ian Foster \\
  University of Chicago\\
  Argonne National Lab \\
  {\texttt{\{chard, foster\}@uchicago.edu}}
}

\date{}

\begin{document}
\maketitle
\begin{abstract}
Relation extraction is an efficient way of mining the extraordinary wealth of human knowledge on the Web. 
Existing methods rely on domain-specific training data or produce noisy outputs.
We focus here on extracting targeted relations from semi-structured web pages given only a short description of the relation.
We present \projTitle{}, an open-domain information extraction method based on a joint graph and language model structure. \projTitle{} can generalize to previously unseen domains without additional data or training and produces only clean extraction results matched to the search keyword. Experiments show that \projTitle{} can improve extraction F1 scores by as much as 34.8\% compared to previous work in a zero-shot domain and zero-shot website setting.
\end{abstract}

\section{Introduction}

Knowledge bases (KBs), such as relational databases and knowledge graphs, are used in many applications to embed human knowledge, recommendation systems~\cite{dong2019building} and digital assistants~\cite{huang2019knowledge,zheng2017natural} are a few examples. 
However, building KBs manually is expensive, so extracting relations from semi-structured pages on the web can be a more efficient method for building extensive KBs~\cite{lockard2020zeroshotceres}. The Internet Movie Database, for example, has over 10 million pages with detailed information on films and TV episodes~\cite{imdb2022stats}. In the science domain, online KBs such as PubMed~\cite{Roberts381}, PolyInfo~\cite{otsuka2011polyinfo} and Polymer Property Predictor and Database~\cite{tchoua2016blending} provide useful knowledge in their respective domains. These pages are ``semi-structured'' because they are automatically populated into an HTML template from a database, and the page structure conveys semantic information. Although the data is often presented in a homogeneous structure within each KB, extracting information from a large number of KBs is still a challenging task because of the heterogeneous structures defined by various KBs. 

Relation extraction is the task of extracting triples of \textit{(subject, predicate, object)} such as \textit{(Polystyrene, glass transition temperature, $100 ^\circ C$)} from texts. RE methods can be classified as either ClosedIE or OpenIE. ClosedIE methods only extract objects for pre-defined predicates, while OpenIE methods extract predicates from the text. However, OpenIE outputs can be noisy due to language ambiguity and variability. The ``schema alignment'' problem, or the difficulty of matching extracted predicates to predefined relations, has limited the use of OpenIE in production. Our goal is to solve schema alignment while extracting tuples.
Existing methods rely on domain-specific training data or produce noisy outputs.
We focus here on extracting targeted relations from semi-structured web pages given only a short description of the relation.
Semi-structured pages are a rich resource of such information since they are most likely populated from a back-end database into a front-end template.
amassed vast amounts of human knowledge as it tightly integrates into every aspect of our life, yet the data exists in a chaotic and heterogeneous manner.
Extracting relations from semi-structured pages poses many challenges. 
Manual curation is laborious and costly. 
Automated methods rely on domain-specific training data or produce noisy outputs that require post-processing to reconcile.
We present \projTitle{}, an open-domain information extraction method based on a joint graph and language model structure. 
\projTitle{} can generalize to previously unseen domains without additional data or training and produces only clean extraction results matched to the search keyword.

Large Language Models (LLMs) have shown exceptional performance on many IE tasks~\cite{devlin2018bert,das2020information,roy2021incorporating}, but they are limited to unstructured text. In this section, we propose a novel OpenIE model, \projTitle{}, that combines the graph model's strength in recognizing structures with the LLM's ability to understand semantics. \projTitle{} takes a target relation keyword and a short textual description of the relation as inputs, and extracts <relation, value> tuples matching the target relation from semi-structured webpages in a zero-shot manner. \projTitle{} can scale to unknown relations without any retraining or finetuning, and it does the extraction and schema alignment in one step to produce clean results.
Experiments show that \projTitle{} can improve extraction F1 scores by as much as 34.8\% compared to previous work in a zero-shot domain and zero-shot website setting.

\section{Problem Definition}
$\mathcal{W}$ is a semi-structured webpage on an entity $\mathcal{T}$ (which we assume has already been identified, for example, from the \texttt{<title>} tag of an HTML webpage).
Given $\mathcal{W}$ and a short textual description $\mathcal{D}$ of a target relation $\mathcal{K}$, extract all relations of $\mathcal{T}$ as tuples $\mathcal{V} = \{ (r_1, v_1), (r_2, v_2), \ldots \}$ from $\mathcal{W}$, 
where $r_i$ is a relation (predicate) that is or is a synonym of $\mathcal{K}$ and $v_i$ is a value (object) corresponding to $r_i$. 

\section{Related Work}
\label{sec:related}
RE on semi-structured data has been widely studied. Early efforts, such as Web Extraction and Integration of Redundant data (WEIR)~\cite{bronzi2013extraction},   focused on rule-based methods.
Solutions involving graph neural networks (GNNs) such as OpenCeres and ZeroshotCeres were proposed as GNNs are well-suited for discovering patterns in semi-structured data~\cite{lockard2020zeroshotceres,lockard2019openceres}, but they did not capture the semantics of the text well.
LLMs were designed to process unstructured text. Previous works have attempted to extend LLMs to semi-structured data~\cite{deng2022dom,aghajanyan2021htlm,xie2021webke}, but they did not explicitly use the structural information and instead relied on the LLMs to learn underlying patterns from semi-structured corpora.




\section{Methodology}
On a semi-structured webpage, the semantics and the structure of the text are equally important indicators of potential relations. Therefore, an effective solution must be capable of handling both parts equally well. 

\begin{figure}
  \centering
  \includegraphics[width=\linewidth]{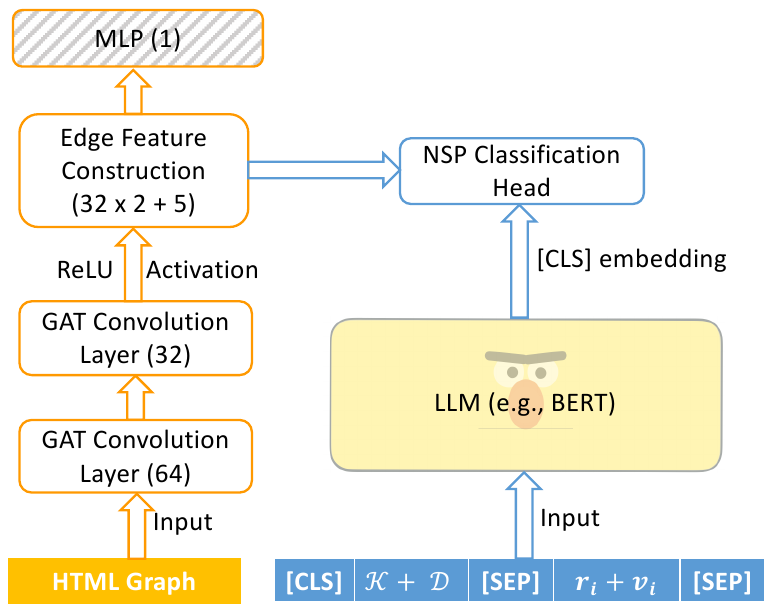}
  \caption{The \projTitle{} model architecture}
  \label{fig:graph_llm_model}
\end{figure}

As shown in Fig.~\ref{fig:graph_llm_model}, \projTitle{} consists of two components: a graph model (left) and an LLM (right). In this paper, the LLM used is ScholarBERT~\cite{hong2023diminishing}, but it can be substituted with any encoder-based Transformer models (such as BERT~\cite{devlin2018bert} or MatBERT~\cite{trewartha2022quantifying}).
The LLM takes the concatenation of $\mathcal{K, D}, r_i, v_i$ as inputs.
It is good at capturing the semantics of the text, but it also needs the layout of the webpage to predict if $(r_i, v_i)$ is a valid pair.
Therefore, we augment the LLM with a graph model, which takes a web page's Document Object Model (DOM) tree as input and extracts DOM features and visual features for every text node and supplies them to the classification head of the LLM.
Text embeddings are not used in the graph model because its message propagation process would dilute and confuse the semantics of the text. Instead, the semantic features are used by the LLM in combination with the graph features to predict whether the extracted value $v$ is valid for the extracted relation and whether the extracted relation matches the search keyword $\mathcal{K}$.

\subsection{Graph Model}
\label{sec:graph}
Extracting <relation, value> node pairs can be viewed as an edge classification problem for graph models.
We pretrained the graph model on classifying whether an edge connects two nodes that make up a true <relation, value> pair. For this purpose, the post-graph convolution edge features are given to a Multi-Layer Preceptron (MLP) for binary classification (Fig.~\ref{fig:graph_llm_model}). After pretraining, the MLP is discarded and the rest of the graph model is frozen. The edge features are provided to the classification head of the LLM as extra input features.

\begin{figure}
  \centering
  \includegraphics[width=\linewidth]{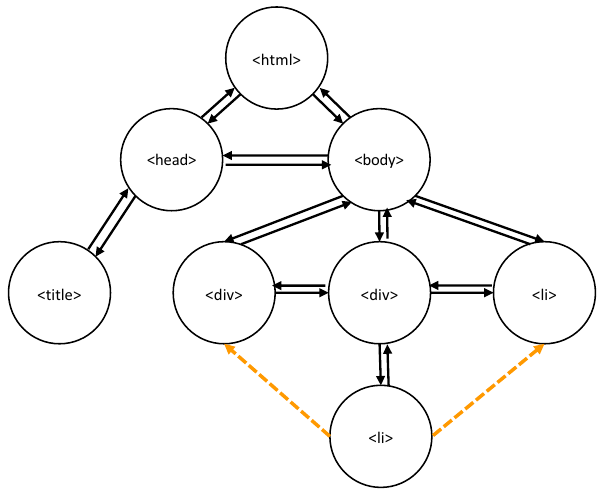}
  \caption{Graph representation of a webpage. DOM node attributes propagate bi-directionally through the black edges but not the dashed orange edges (``virtual edges''), which are only used for classification.}
  \label{fig:edge}
\end{figure}

To construct a graph for a webpage, each DOM node is represented as a node in the graph with 3 features:
\begin{itemize}
    \item \texttt{tag}: The one-hot encoded HTML tag of the node (``div'', ``li'', etc.);
    \item  \texttt{index}: The index of the current node among its siblings;
    \item  \texttt{text\_freq}: the average number of occurrences of the node's text per page within the website.
\end{itemize}

Edges connect each node to its parent, children, and immediate siblings (Fig.~\ref{fig:edge}).
Some relation-value pairs may exist in other types of connections. Adding all possible pairwise edges would be inefficient, so we only add a \textit{virtual edge} between two nodes ($K$ and $V$) if and only if all four conditions are met: 
\begin{itemize}
    \item  they are within $h$ hops of each other;
    \item  both contains text with length in $[l_{min}, l_{max}]$ (to filter out text too long or too short to be a relation or value);
    \item  the \texttt{text\_freq} of $K$ $>m$ (as $r_i$ tends to be common on semi-detailed pages of the same website);
    \item  the \texttt{text\_freq} of $V$ $<n$ (as $v_i$ would likely vary from page to page).
\end{itemize}

The virtual edges are hidden from the graph model's message-passing process to avoid skewing the structure of the webpage.
Edge features include: 
\begin{itemize}
    \item Node features of $K$ and $V$;
    \item Distance (number of hops) between $K$ and $V$;
    \item Differences in the horizontal and vertical coordinates between $K$ and $V$;
    \item Differences in the heights and widths of the bounding boxes of $K$ and $V$.
The features are updated post-graph convolution for every edge including virtual edges.
\end{itemize}

\subsection{The Language Model}
We adapted Next Sequence Prediction (NSP) to solve the relation extraction and schema alignment problem in one step.
In our approach, NSP is used to predict whether two sentences, one consisting of the search keyword $\mathcal{K}$ and the description $\mathcal{D}$ (Sentence A) and the other consisting of text in nodes $K$ and $V$ connected by an edge in the previous step (Sentence B), form a true relation-value pair based on both the text features (from the LLM) and the edge features of the edge connecting $K$ and $V$ (from the graph model).

\section{Experimental Evaluation}
\projTitle{} is designed for scalability so it is tested in a zero-shot setting on previously unseen websites. We compare its performance to a heuristic baseline and the methods introduced in Section~\ref{sec:related} where available and test its ability to extract relations from websites within the same vertical, transfer knowledge from one vertical to another, and examine the impact of training data and different model components on overall performance.
To evaluate our model, we conducted experiments using three publicly available datasets, Structured Web Data Extraction (SWDE)~\cite{hao2011one} and expanded SWDE~\cite{lockard2020zeroshotceres}, and PPPDB~\cite{tchoua2016blending}  (Table~\ref{tab:datasets}). 
We compare \projTitle{}'s performance to a heuristic baseline and the methods introduced in Section~\ref{sec:related} where available.
The experiments were conducted on a cloud computing provider for $\sim20$ GPU-hours with one V100 GPU.

\begin{table*}[htbp]
\centering
\caption{Statistics of the SWDE, Retail, and Expanded SWDE datasets used in the experiments.}
\label{tab:datasets}
\resizebox{0.7\linewidth}{!}
{
\begin{tabular}{ccccc}
\hline
\textbf{Dataset} & \textbf{Verticals} & \makecell{\textbf{Websites} \\ \textbf{per Vertical}} & \makecell{\textbf{Webpages} \\ \textbf{per Website}} & \makecell{\textbf{Annotated} \\ \textbf{Relations/Website}}\\
\hline
SWDE             & 4                  & 10                         & $\sim$200-2000            & 3-5                                  \\
Expanded SWDE    & 3                  & 10                         & $\sim$400-2000            & 5-30      \\
PPPDB           & 1                  & 2                          & $\sim$25          & 3-5   \\
\hline
\end{tabular}
}
\end{table*}

\begin{table*}[htbp]
\centering
\caption{Examples of search keywords and descriptions used as inputs. The ``Ground Truth Relations'' column lists the expressions used by different websites to represent the same relation. The search keywords and descriptions are manually provided. The search keywords are usually synonyms or hypernyms of the ground truth relations.}
\label{tab:search_key}
\resizebox{\linewidth}{!}
{
\begin{tabular}{ccc}
\hline
\textbf{Search Keyword} & \textbf{Short Description}                                                                                   & \textbf{Ground Truth Relations}   \\
\hline
fuel economy            & The fuel economy of an automobile relates distance traveled by a vehicle and the amount of fuel consumed.    & MPG, Gas Milage, EPA Fuel Economy \\
price                   & The price of is the amount of money expected, required, or given in payment for something.                   & Price, MSRP, Starting MSRP        \\
finish type & The finish is the surface appearance of a manufactured material or object & Surface Type, Texture/finish\\
compatibility & Compatibility is the ability of one computer, piece of software, etc. to work with another& App Compatible, Compatible Platform(s)\\ 
glass transition temperature & The glass transition temperature is the temperature at which a material transitions from a hard state into a rubbery state  & Tg(K)\\
Flory-Huggins parameter & The Flory-Huggins interaction parameter quantifies the interaction between different types of molecules in a polymer mixture& $\chi$, $\chi^N$\\

\hline
\end{tabular}
}
\end{table*}

\subsection{Intra-vertical Extraction}

The intra-vertical experiments replicated the scenario where prior knowledge of the entities and relations in the target vertical was available and could be used to train the model in order to extract new information from a previously unseen website in the same vertical. 
In each vertical, we randomly selected a number of websites of which the total number of webpages are around 50\% of all webpages in said vertical and used these websites as training data, while the remainder made up the test set. 
For SWDE and Extended SWDE, the Graph model is randomly initialized and the LLM is the pre-trained Bert-base model.
For the PPPDB dataset, the Graph model inherits the pre-trained weights from SWDE since  the sample size is too small to train the graph model from scratch. ScholarBERT-100 is selected as the LLM model due to its science-focused pretraining.

Table~\ref{tab:intra_swde}, \ref{tab:intra_exp_swde}, \ref{tab:intra_pppdb} shows the intra-vertical extraction performance on the SWDE, expanded SWDE, and PPPDB datasets respectively. 
For each dataset, the performance of the \projTitle{} model was compared to the baseline as well as existing methods (including SOTA) from literature where available.
\projTitle{} was able to achieve the highest F-1 score on average (0.72) across all datasets. On the expanded SWDE dataset it outperformed previous SOTA methods by as much as 22.4\% as shown in Table~\ref{tab:intra_exp_swde}. 

\begin{table*}
\centering
\caption{Intra- and inter-vertical OpenIE extraction performances on the SWDE dataset.}
\label{tab:intra_swde}
\resizebox{\linewidth}{!}
{
\begin{tabular}{c|c|ccc|ccc|ccc|ccc|c}
\hline
        &              & \multicolumn{3}{c|}{\textbf{Auto}} & \multicolumn{3}{c|}{\textbf{Book}} & \multicolumn{3}{c|}{\textbf{Camera}} & \multicolumn{3}{c|}{\textbf{Movie}} & 
                    \multicolumn{1}{c}{\textbf{Average}}\\
Method     & Vertical        & Precision  & Recall  & F-1 Score  & Precision  & Recall  & F-1 Score  & Precision   & Recall   & F-1 Score  & Precision   & Recall  & F-1 Score & F-1 Score \\
\hline
Heuristic Baseline & - & 0.07       & 0.19    & 0.10       & 0.12       & 0.18    & 0.14       & 0.27        & 0.52     & 0.36       & 0.13        & 0.22    & 0.16       & 0.19 \\
\projTitle{}  & Intra      & 0.65 & \textbf{1.00} & 0.78 & \textbf{0.92} & \textbf{0.70} & \textbf{0.79} & \textbf{0.95} & \textbf{0.96} & \textbf{0.96} & 0.70 & \textbf{1.00} & 0.83 &\textbf{0.84}\\
\projTitle{}  & Inter      & \textbf{0.66} & \textbf{1.00} & \textbf{0.79} & 0.90 & 0.52 & 0.66 & 0.79 & 0.78 & 0.79 & \textbf{0.96} & 0.85 & \textbf{0.90} & 0.79\\
\hline
\end{tabular}
}
\end{table*}

\begin{table*}
\centering
\caption{Intra- and inter vertical OpenIE extraction performances on the expanded SWDE dataset.}
\label{tab:intra_exp_swde}
\resizebox{\linewidth}{!}
{
\begin{tabular}{c|c|ccc|ccc|ccc|c}
\hline
        &  & \multicolumn{3}{c|}{\textbf{Movie}}            & \multicolumn{3}{c|}{\textbf{NBA Player}}       & \multicolumn{3}{c|}{\textbf{University}}       & 
          \multicolumn{1}{c}{\textbf{Average}}\\
Method  & Vertical  & Precision     & Recall        & F-1 Score     & Precision     & Recall        & F-1 Score     & Precision     & Recall        & F-1 Score    & F-1 Score \\
\hline
Heuristic Baseline & -    & 0.23          & 0.16          & 0.14          & 0.19          & 0.37          & 0.25          & 0.15          & 0.11          & 0.13          & 0.17\\
WEIR & Intra      & 0.14          & 0.1           & 0.12          & 0.08          & 0.17          & 0.11          & 0.13          & 0.18          & 0.15         & 0.13 \\
OpenCeres & Intra & \textbf{0.71}          & 0.84          & 0.77          & \textbf{0.74}          & 0.48          & 0.58          & \textbf{0.65}          & 0.29          & 0.4          & 0.58 \\
ZSCeres & Intra & 0.49        & 0.51          & 0.50          & 0.47        & 0.39          & 0.42          & 0.50          & 0.49          & 0.50          & 0.47 \\
WebKE   & Intra  & -             & -             & 0.76          & -             & -             & 0.64          & -             & -             & 0.34          & 0.58\\
ZSCeres & Inter & 0.43        & 0.42          & 0.42          & 0.48        & 0.49          & 0.48          & 0.49          & 0.45          & 0.47          & 0.46\\

\projTitle{}  & Intra & \textbf{0.71} & \textbf{0.92} & \textbf{0.80} & 0.69 & \textbf{0.79} & \textbf{0.74} & 0.52 & \textbf{0.67} & \textbf{0.59} & \textbf{0.71}\\

\projTitle{}  & Inter & 0.65 & 0.88 & 0.75 & 0.52 & 0.71 & 0.60 & 0.44 & 0.61 & 0.51 & 0.62\\
\hline
\end{tabular}
}
\end{table*}

\begin{table*}
\centering
\caption{Intra- and inter-vertical OpenIE extraction performances on the PPPDB dataset.}
\label{tab:intra_pppdb}
\resizebox{0.8\linewidth}{!}
{
\begin{tabular}{c|c|ccc|ccc|ccc|c}
\hline
          &          & \multicolumn{3}{c|}{\textbf{Glass Transition Temperature}} & \multicolumn{3}{c|}{\textbf{Flory-Huggins parameter}} & \multicolumn{1}{c}{\textbf{Average}} \\
Method    & Vertical  & Precision  & Recall  & F-1 Score  & Precision  & Recall  & F-1 Score    & F-1 Score  \\
\hline
Heuristic Baseline & - & 0.00 & 0.00 & 0.00 & 0.00 & 0.00 & 0.00 & 0.00\\
\projTitle{} & Intra & \textbf{0.87} & \textbf{0.55} & \textbf{0.67} & \textbf{0.68} & \textbf{0.49} & \textbf{0.57} & \textbf{0.62} \\
\projTitle{} & Inter & 0.82 & \textbf{0.55} & 0.66 & 0.61 & 0.48 & 0.54 &   0.60\\
\hline
\end{tabular}
}
\end{table*}

\begin{table*}[htbp]
\centering
\caption{The top (green) and bottom (orange) 3 relations in SWDE and Expanded SWDE datasets based on inter-vertical extraction performance of the \projTitle{} model.}
\label{tab:per_attr}
\resizebox{0.8\linewidth}{!}
{
\begin{tabular}{cccc|cccc}
\hline
 \multicolumn{4}{c|}{\textbf{SWDE}}  & \multicolumn{4}{c}{\textbf{Expanded SWDE}} \\
 Search Keyword                       & Precision          & Recall          & F-1 Score          & 
 Search Keyword                       & Precision          & Recall          & F-1 Score          \\
\hline
{\color[HTML]{0F936A} engine}         & 1.0                & 1.0             & 1.0                & 
{\color[HTML]{0F936A} MPAA rating}    & 0.84               & 0.96            & 0.90               \\
{\color[HTML]{0F936A} fuel economy}   & 0.99               & 1.0             & 1.0                &
{\color[HTML]{0F936A} salary}         & 0.88               & 0.90            & 0.89               \\
{\color[HTML]{0F936A} director}       & 1.0                & 0.94            & 0.97               &
{\color[HTML]{0F936A} highest degree} & 0.82               & 0.76            & 0.79               \\
{\color[HTML]{CE541A} ISBN13}         & 0.71               & 0.70            & 0.70               &
{\color[HTML]{CE541A} keyword}        & 0.36               & 0.65            & 0.46              \\
{\color[HTML]{CE541A} model}          & 0.30               & 0.99            & 0.46               &
{\color[HTML]{CE541A} tuition}        & 0.34               & 0.55            & 0.42               \\
{\color[HTML]{CE541A} price}          & 0.24               & 1.0             & 0.39               &
{\color[HTML]{CE541A} room and board} & 0.19               & 0.76            & 0.30               \\
\hline
\end{tabular}
}
\end{table*}

\begin{table*}[htbp]
\centering
\caption{Ablation results on the SWDE dataset.}
\label{tab:ablation}
\resizebox{\linewidth}{!}
{
\begin{tabular}{c|ccc|ccc|ccc|ccc|c}
\hline
                    & \multicolumn{3}{c|}{\textbf{Auto}} & \multicolumn{3}{c|}{\textbf{Book}} & \multicolumn{3}{c|}{\textbf{Camera}} & \multicolumn{3}{c|}{\textbf{Movie}} & \textbf{Average}\\
Method             & Precision  & Recall  & F-1 Score  & Precision  & Recall  & F-1 Score  & Precision   & Recall   & F-1 Score  & Precision   & Recall  & F-1 Score & F-1 Score \\
\hline
\projTitle{} & 0.66 & 1.00 & 0.79 & 0.9 & 0.52 & 0.66 & 0.79 & 0.78 & 0.79 & 0.96 & 0.85 & 0.9 & 0.79\\
\hline
\multirow{2}{*}{- Graph Features}        & 0.54    & 1.00   & 0.70    & 0.91    & 0.41    & 0.57    & 0.94    & 0.60    & 0.73    & 0.91    & 0.91   & 0.91 &  0.73  \\
                                         & (-0.12) & (0.00)   & (-0.09)  & (+0.01)  & (-0.11) & (-0.09) & (+0.15) & (-0.18) & (-0.06) & (-0.05)  & (+0.06) & (+0.01) & (-0.06) \\
\hline                                         
\multirow{2}{*}{- 1/3 training examples} & 0.79    & 0.65    & 0.71    & 0.83    & 0.30    & 0.44    & 0.42    & 0.61    & 0.50    & 0.70    & 0.92   & 0.79 &   0.61   \\
                                         & (+0.13) & (-0.35) & (-0.08)  & (-0.07)  & (-0.22) & (-0.22) & (-0.37) & (-0.17) & (-0.29) & (-0.26) & (+0.07) & (-0.11) & (-0.18) \\
\hline                                         
\multirow{2}{*}{- 2/3 training examples} & 0.54    & 1.00   & 0.70    & 0.94    & 0.24    & 0.38    & 0.67    & 0.24    & 0.35    & 0.63    & 0.98   & 0.76    & 0.55    \\
                                         & (-0.11) & (0.00)   & (-0.09)  & (+0.04)  & (-0.28) & (-0.28) & (-0.12)  & (-0.54) & (-0.44) & (-0.36) & (+0.07) & (-0.14) & (-0.24)        \\
\hline
\end{tabular}
}
\end{table*}

\subsection{Inter-vertical Extraction}
The intra-vertical experiments showed promising performance, but for scalability, we do not want to build models for every vertical. Moreover, oftentimes labeled data was not available in the desired vertical. So transferring knowledge learnt from one vertical to another is vital for scalability. In this section, the experiments were conducted in a K-fold fashion. The model was trained on all verticals except the one currently used for testing. 
For the inter-vertical experiments, we trained \projTitle{} on all verticals except the one being tested. \projTitle{} performed the best, with an average F-1 score of 0.66 across all datasets (Table~\ref{tab:intra_swde}, \ref{tab:intra_exp_swde}), \ref{tab:intra_pppdb}. On the expanded SWDE, its F-1 score exceeded previous methods by 34.8\%. \projTitle{}'s average F-1 was 5.33 points higher in the intra-vertical tests than in the inter-vertical tests on the same datasets, as expected since the webpages in the intra-vertical training data were more similar to the test pages. The PPPDB dataset has seen the smallest gap between intra- and inter-vertical tests because the webpages in its verticals (Glass Transition Temperature and Flory-Huggins parameter) are the most similar compared to the webpages from different verticals in SWDE or Expanded SWDE.

\subsection{Error Analysis}

Table~\ref{tab:per_attr} shows the three best and three worst target relations in terms of F-1 scores for the SWDE and Extended SWDE datasets, from which we make three observations.

(i) The model tends to perform better 
when there is a strong semantic relationship between the object and the predicate. E.g., the values for the relation ``engine'' often contains ``horsepower'' or ``turbo''. Such words are uncommon in other relations, so the model can easily match them to their corresponding relation.

(ii) The model struggles with relations with broad potential values. E.g., the values for ``model'' and ``ISBN13'' are alphanumeric strings; relations such as ``keyword'' can have numerous possible values. In such cases the model cannot eliminate false positives effectively by semantics, leading to low precision.

(iii) The amount of noise on a webpage can significantly affect the performance. While \projTitle{} had a low F-1 score on ``price'' from the camera vertical in SWDE, it had little problem extracting ``salary'' for NBA players in expanded SWDE. Both values were dollar amounts, but webpages about cameras may have many other dollar values such as MSRP and discount, whereas on a page about an NBA player, the only dollar value present around the ``salary'' node is the salary amount.

\subsection{Ablation}

Table~\ref{tab:ablation} shows the changes in extraction performance when the graph features were removed and when reducing the amount of data used to pretrain the graph model. In every ablation case, the average F-1 score on the SWDE dataset dropped by 6 to 24 points.
Interestingly, when the graph component was pretrained on less data, the model performed worse than having no graph component at all.
Even without the graph features, the LLM could recognize a lot of false inputs based on the semantics only. As an example, for the target relation \textit{``price''}, the LLM could predict (``MSRP'', ``Charles Dickens'') and (``author'', ``Charles Dickens'') were not valid extractions due to the mismatched semantics.
Nevertheless, an under-trained graph component could provide noise, rather than helpful features, to the language model, thus interfering with the language model's ability to make the correct final prediction.

\subsection{Summary}
\projTitle{} is an OpenIE model for extracting relations from the Web. We framed the task of extraction relations from webpages as a Next Sentence Prediction problem for a language model, which we augmented with a graph model to extract DOM features from webpages. Experimental results showed that \projTitle{} outperformed previous work by as much as 34.8\% F1.

\bibliographystyle{acl}
\bibliography{refs}

\end{document}